\documentclass{vgtc}                          

\ifpdf
  \pdfoutput=1\relax                   
  \usepackage{graphicx}                
  \DeclareGraphicsExtensions{.pdf,.png,.jpg,.jpeg} 
\else
  \usepackage{graphicx}                
  \DeclareGraphicsExtensions{.eps}     
\fi%

\graphicspath{{figures/}{pictures/}{images/}{./}} 

\usepackage{microtype}                 
\PassOptionsToPackage{warn}{textcomp}  
\usepackage{textcomp}                  
\usepackage{mathptmx}                  
\usepackage{times}                     
\usepackage{cite}                      
\usepackage{tabu}                      
\usepackage{booktabs}                  
\usepackage{soul}
\usepackage{comment}
\usepackage{xcolor}
\usepackage{academicons}
\usepackage{tikz} 
\definecolor{lime}{HTML}{A6CE39}
\usepackage{orcidlink}
\usepackage{wasysym}
\usepackage{multirow}
\PassOptionsToPackage{hyphens}{url}\usepackage{hyperref}

\usepackage{svg}
\usepackage{expdlist}
\usepackage{marvosym}
\usepackage{pdfpages}
\usepackage{ amssymb }
\usepackage{flushend}
\usepackage{sidenotes}
\usepackage{xspace}
\newcommand{\ie}{i.\,e.\@\xspace}
\newcommand{\eg}{e.\,g.\@\xspace}
\newcommand{\etal}{and colleagues\@\xspace}

\newcommand{\inlinevis}[3]{\raisebox{#1}[0pt][0pt]{\includegraphics[height=#2]{#3}}}

\newcommand{\todo}[1]{{\color{red}\bf{TODO: #1}\normalfont}}

\newcommand{\change}[1]{{\color{black}{#1}\normalfont}}

\newlength{\picturewidth}

\onlineid{7520}

\vgtccategory{Research}

\vgtcinsertpkg

\title{Open Questions about the Visualization of Sociodemographic Data}
\author{
\authororcid{Florent Cabric}{0000-0002-9326-9441}\thanks{e-mail: florent.cabric@inria.fr}\\ %
         \parbox{1.2in}{\scriptsize \centering LISN, Université Paris-Saclay, CNRS,\\ Inria, Saclay, France} %
\and \authororcid{Margrét Vilborg Bjarnadóttir}{0000-0003-2955-1992}\thanks{e-mail: mbjarnad@umd.edu}\\ %
     \parbox{1.4in}{\scriptsize \centering DOIT, Robert H. Smith School of Business, University of Maryland, United States}
\and \authororcid{Anne-Flore Cabouat}{0000-0002-3327-2729}\thanks{e-mail: anne-flore.cabouat@inria.fr}\\ %
     \parbox{1.2in}{\scriptsize \centering LISN, Université Paris-Saclay, CNRS,\\ Inria, Saclay, France}%
\and \authororcid{Petra Isenberg}{0000-0002-2948-6417} \thanks{e-mail: petra.isenberg@inria.fr}\\ %
     \parbox{1.2in}{\scriptsize \centering LISN, Université Paris-Saclay, CNRS,\\ Inria, Saclay, France}
     }
\abstract{
This paper collects a set of open research questions on how to visualize sociodemographic data. Sociodemographic data is a common part of datasets related to people, including institutional censuses, health data systems, and human-resources files. This data is sensitive, and its collection, sharing, and analysis require careful consideration. For instance, the European Union, through the General Data Protection Regulation (GDPR), protects the collection and processing of any personal data, including sexual orientation, ethnicity, and religion. Data visualization of sociodemographic data can reinforce stereotypes, marginalize groups, and lead to biased decision-making. It is, therefore, critical that these visualizations are created based on good, equitable design principles. In this paper, we discuss and provide a set of open research questions around the visualization of sociodemographic data. Our work contributes to an ongoing reflection on representing data about people and highlights some important future research directions for the VIS community. A version of this paper and its figures are available online at \href{https://osf.io/a2u9c/}{osf.io/a2u9c}.
} 

\CCScatlist{
  \CCScatTwelve{Human-centered computing}{Visu\-al\-iza\-tion}{}{};
}

\begin{document}


\firstsection{Introduction}
\maketitle
Data visualizations are commonly based on an abstraction of data according to some criteria, such as an average or a clustering. These abstractions enable us to make comparisons,3 for instance, between multiple groups or to identify outliers that do not match the characteristics of the group. When we visualize data about people, this kind of abstraction may lead to the promotion of stereotypes and biases. Stereotypes 
\change{are built on} grouping together people with 
presumed shared characteristics \change{as a means} to describe group membership, contrast that group with others, and identify outsiders \cite{Becker:1963:Outsiders,Nelson:2009:Stereotype,Heywood:2022:Stereotype}. Abstraction can also lead 
to invisibilizing marginalized groups \change{and their issues \cite{Schwabish:2021:Harm,Rall2016Invisible,Herzog:2018:Invisibilization}}. \change{Thus, the process of abstraction of data in visualizations can be very harmful --- for example, by invisibilizing marginalized groups, leading to biased decisions, or conveying a feeling of not being included.} 
\change{The design of nonharmful visualization raises many design \cite{Schwabish:2021:Harm} and ethical \cite{Correll:2019:Ethical} questions.}

Our work is in line with previous work on 
how to visualize data about people \cite{Correll:2019:Ethical,Dork:2013:Political,Dhawka:2023:Race,Schwabish:2021:Harm}. In particular, we focus on expanding on the ``Do No Harm Guide'' \cite{Schwabish:2021:Harm} that discusses a set of considerations for designing visualizations considering diversity, equity, and inclusion. We elaborate on this work through a discussion of open research questions. Our work is a reflection on the challenges that researchers in the community need to tackle to effectively and equitably visualize sociodemographic data \change{in a nonharmful way} and an opportunity to discuss our current practices. 

\section{Protection and Visualization of Sociodemographic Data}

The protection of private attributes has a long history but has seen rapid changes in recent decades.
The number of policies regarding the protection of sociodemographic data is growing across companies, states, and institutions. One such policy, the EU's General Data Protection Regulation (GDPR) \cite{GDPR:2016}, strongly regulates the processing of five categories of sensitive data: \emph{personal, trade-union membership, genetic or biometric, health-related,} and \emph{sex life or sexual orientation}. 
The UK \cite{UK:2010:EqualityAct} defines nine attributes as protected: \emph{age, disability, sex, sexual orientation, gender reassignment, marriage and civil partnerships, pregnancy/maternity, race,} and \emph{religion/belief}. Discrimination against people based on protected attributes is forbidden by law. Similarly, Canada differentiates 16 characteristics about people \cite{Canada:RightACT} that require protection against discrimination: \emph{race, national or ethnic origin, color, religion, age, sex, sexual orientation, gender identity and expression, marital status, family status, genetic characteristics, disability, creed, irrational fears} and \emph{source of income}; and the US defines 11 \cite{USA:CivilRight}: \emph{race, religion, national origin, age, sex, sexual orientation, pregnancy, family status, disability status, veteran status,} and \emph{genetic information}.

Several papers have discussed broader issues around the visualization of data about people. The belief that data visualizations might be neutral has been disparaged by many: Visualizations are political \cite{Dork:2013:Political}, can promote empathy \cite{Boy:2017:Antrhopographics}, often do not surface inherent uncertainty \cite{Hullman:2020:WhyNotUncertainty}, and have even been called inhumane \cite{Dragga:2001:Inhumanity}. Beyond the visualization, data and its collection and processing are also regularly prone to biases. Criado Perez \cite{CriadoPerez:2019:Invisible} has summarized how unrepresentative data collection about people (for example, only collecting data from men) has led to overgeneralizations to the 
general population and thus has caused discrimination and other problems such as discomfort in women's daily lives \cite[Chapter 8]{CriadoPerez:2019:Invisible} or medical complications \cite[Chapter 10]{CriadoPerez:2019:Invisible}. 

Out of the above-mentioned personal attributes, a few have been the subject of dedicated papers on how to visualize them in a nonharmful way. Disability has gained attention from visualization researchers, with most work focused on making data visualization accessible to people with disabilities \cite{Wu:2021:Accessibility,Marriott:2021:Inclusive,Elmqvist:2023:Blind,Elavsky:2022:Accessibility}. Only a few studies have explored how to represent data about people with disabilities. For example, Barstow and colleagues \cite{Barstow:2019:Symbols} investigated the design of inclusive visual symbols beyond the ``traditional'' wheelchair icon, which fails to represent many people with a physical disability who need specific access or consideration. 
Race or ethnicity has received attention regarding the design of nonharmful visualization. 
Schwabish and Feng \cite{Schwabish:2021:Harm} discussed a set of considerations for data collection, use of language, visibility of data, and use of colors, icons, and shapes. The authors emphasize talking to people to ensure they are represented correctly and harmlessly 
in data visualizations. Dhawka and colleagues \cite{Dhawka:2023:Race} explored the challenges and opportunities of representing diverse people grouped by race using anthropographics. Gender has also been the subject of studies related to the diversity of the community itself \cite{Tovanich:2022:gender}. Papers on the visual representation of gender as a demographic variable are rare. In a blog post, Muth \cite{Muth:2018:alternative} collected multiple examples of professional visualizations of gender by institutions such as Bloomberg and the New York Times. 
The visualizations used a variety of colors, including pink/blue, but also many other combinations, such as orange/green 
or green/purple, without a clear consistency. Finding the right visual variables, layouts, and 
marks has been shown to be challenging for these sociodemographic attributes (disability, ethnicity, and gender), and many others may surface as more of the protected attributes are 
explored. However, multiple personal attributes raise common open research questions that will be discussed next. 


%
\vspace{-.2em}
\section{Some Open Questions about the Visualization of Sociodemographic Data}
As pointed out in the introduction, 
visualizations are often inherently built on a process of abstraction. When data about people have been abstracted into groups, these groups need to be assigned marks, visual variables, and a layout. This visualization design process involves choices that may reflect a certain way of thinking or ``not thinking'' \cite[Preface]{CriadoPerez:2019:Invisible}, built-in default encodings, or the use of over-general guidelines. Visualization challenges in this process are related to the potential amplification of stereotypes and the marginalization of groups of people. In addition, designers have the difficult task of balancing how the visualized populations will see themselves reflected in the visualization and the ability of decision-makers to use the visualization effectively. We next outline open questions we would like to discuss as part of the Vis4Good workshop.

\vspace{-.5em}
\subsection{Balancing \change{Efficiency}}

\setlength{\picturewidth}{\columnwidth}
\begin{figure}[!b]
 \centering 
 \vspace{-.8em}
\includegraphics[width=0.95\columnwidth]{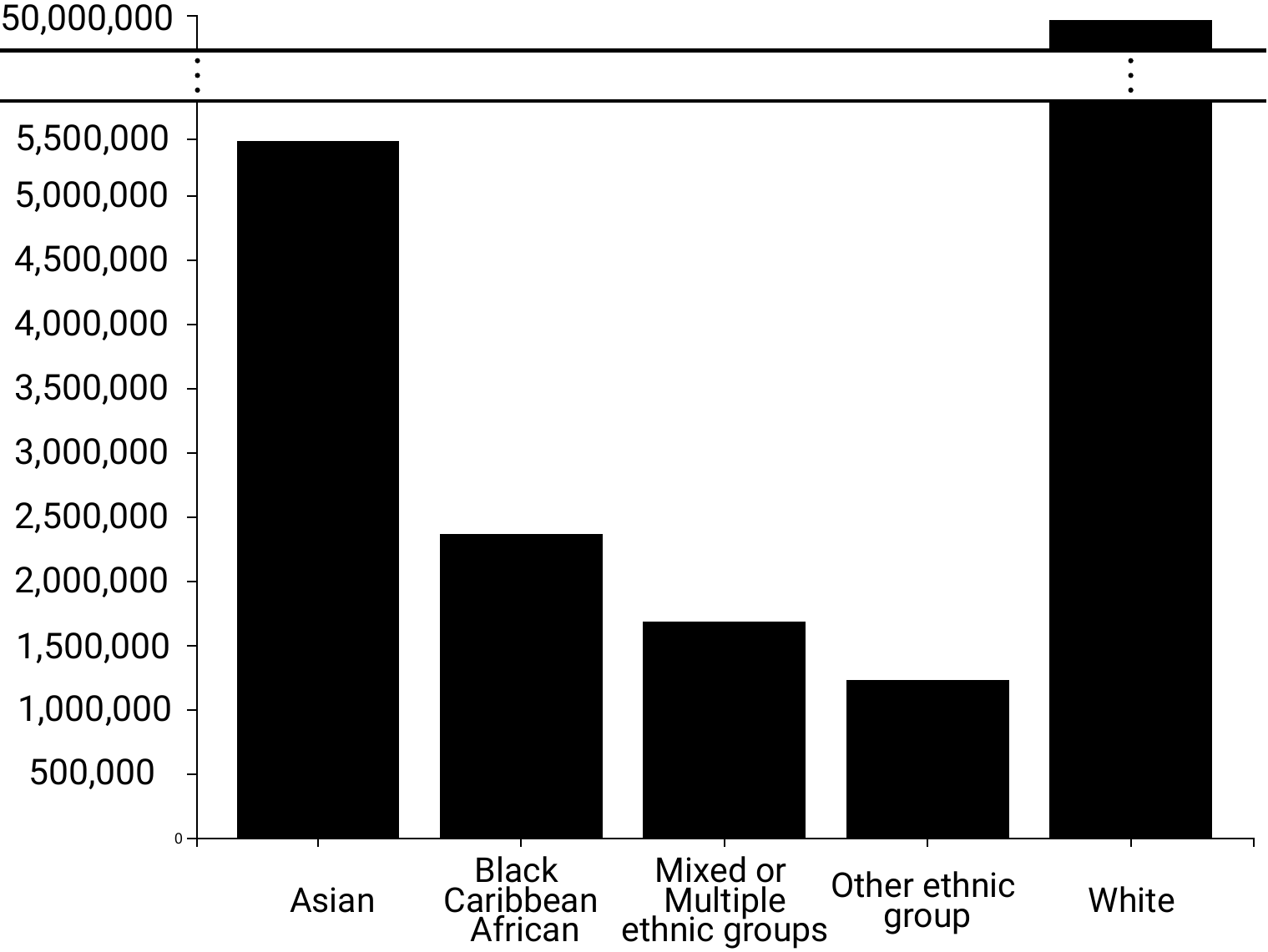}
\vspace{-.2em}
\caption{England's and Wales's residents according to five different ethnic groups. Contrast this figure with \autoref{fig:twentygroups} and consider the balance between information and visualization simplicity and how included different types of viewers may feel. The data is from the UK census and uses its data labels \cite{ONS:2022:census}.
\textit{Note: the y-axis is broken between the  5.5M and 50M ticks}.}
\label{fig:fivegroups}
\end{figure}
 
Visual decision-making tasks are cognitively complex: The reader must understand the visualization, define the underlying decision problem, compare different options, and then make one choice \cite{Dimara:2022:Decisions}.
Such complexity may require balancing inclusive visual design with the need for \change{time-efficient} task completion. Air traffic control, crowd management, or emergency medicine regularly require decision-makers to complete critical tasks quickly and accurately. To reduce the time taken to make a decision, people can rely on stereotypes to activate heuristics \cite{Bodenhausen:1985:Stereotypes,Bodenhausen:1990:Stereotypes}, sometimes at the cost of accuracy \cite{Payne:1988:DM}. This Speed-Accuracy Trade-off (SAT) \cite{Larson:2022:SAT} has been discussed for a long time in psychology \cite{Gigerenzer:2011:Heuristics}. 
In the context of visualization, we are interested in discussing the impact of stereotyped visual representations on the decision-making processes.

Creating visualization with stereotyped attributes such as red and blue colors for gender may speed up cognitive tasks and decisions. Studies have assessed the effectiveness of using semantically resonant colors \cite{Lin:2013:Semantic}. For example, using a color associated with the data represented, such as 
yellow for a bar representing bananas. In contrast, there is some evidence that using a reversed mapping --- associating a visual mapping that is typically associated with a specific group (brighter colors = women) to another group (brighter colors = men) --- slows down classification and association tasks \cite{semin:brightness:2014}. 

We are not aware of studies that evaluated the \change{efficiency} of stereotyped and unstereotyped visualizations for common tasks \cite{Brehmer:2013:Tasks}. Suppose previous results on semantically resonant colors are confirmed with stereotyped visualizations. In that case, a new dilemma will arise for designers of visualizations used in situations that require quick and accurate task completion: Is the \change{efficiency} of my visualizations worth the use of stereotyped data encodings? One may argue that this question is not even worth asking 
because the well-being of the people represented should trump concerns about \change{efficiency}. Yet, people's well-being might also depend on quick and correct decisions. What has not been studied to date is: To what extent do the potential benefits of stereotyped visualizations matter? 
Perhaps the \change{efficiency} benefits are so small that we should never consider using stereotyped visualizations, and research should focus on raising awareness of stereotypes. Or, perhaps the benefits are significant enough to motivate future works on how to avoid stereotyped visualizations while still achieving the same \change{efficiency}. For example, training people to associate groups with non-stereotyped visual variables such as colors may solve the problem; but how and to what extent to invest in such training in different scenarios is an open question. To be clear, we do not call for the use of stereotyped visual encodings. Instead, we call for more research to fully understand the effects of using undesired stereotyped visualizations.
\vspace{-0.2em}
\subsection{Balancing Simplicity}

\setlength{\picturewidth}{\columnwidth}
\begin{figure}[!b]
 \centering 
 \vspace{-1.5em}
\includegraphics[width=0.95\columnwidth]{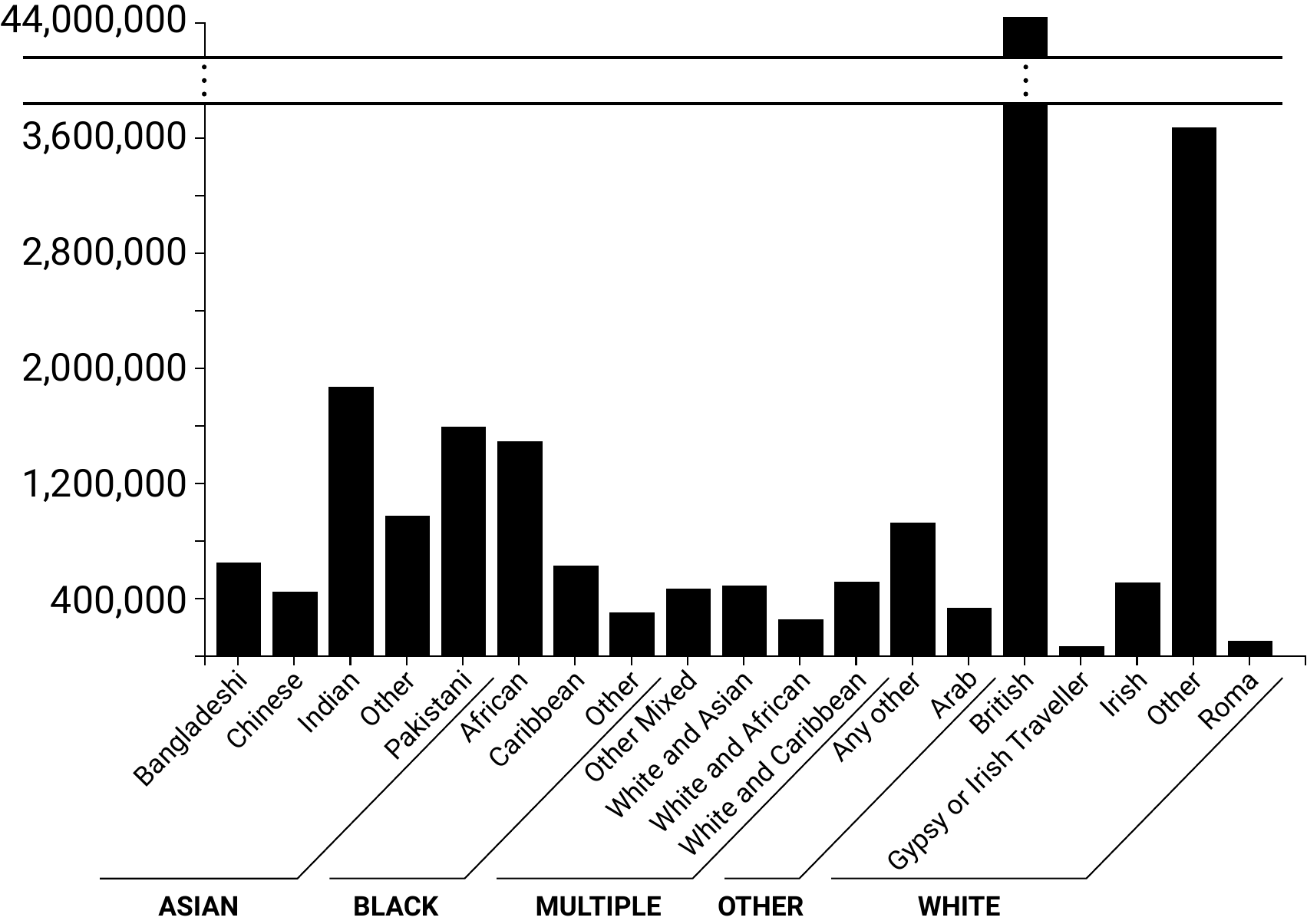}
  \vspace{-.2em}
 \caption{England's and Wales's residents grouped into 20 different ethnic groups. Contrast this figure with \autoref{fig:fivegroups} and consider the balance between information and visualization simplicity and how included different types of viewers may feel. The data is from the UK census and uses its data labels \cite{ONS:2022:census}. \textit{Note: the y-axis is broken between the 3.6M and 40M ticks}. }
 \label{fig:twentygroups}
\end{figure}
In their ``Do No Harm Guide,'' Schwabish and Feng \cite{Schwabish:2021:Harm} recommend that ``data visualizations should use more complex designs if
that would more accurately reflect and promote a better understanding of the topic being shown.'' \change{In addition, Baumer \etal \cite{Baumer:2022:OfCourseItsPolitical} argue that complexity can lead to inclusivity and help data visualization readers to gain deeper insights.}
We relate complexity 
here to the number of data categories being displayed. To facilitate data exploration, designers must disaggregate and split specific data dimensions into different factors and subfactors. For instance, the 2021 UK Census dataset \cite{ONS:2022:census} proposed three levels of disaggregation for ethnicity data: 5 (see \autoref{fig:fivegroups}), 20 (see \autoref{fig:twentygroups}), and 288 categories. 

On the one hand, the representation of a small number of groups can result in the invisibilization of small and sometimes marginalized groups. Invisibilizing groups can reinforce the feeling of being outside society for the people represented \cite{Herzog:2018:Invisibilization}. On the other hand, \textit{hypervisibilizing} groups \cite{Settles:2019:Hypervisibilize} can be interpreted as \textit{tokenization} of the people represented \cite{Wingfield:2014:Token} and lead to increased surveillance from the main group. Overrepresentation in the data or visualization can 
reinforce feelings of marginalization. 

The continuum of disaggregation is vast, and visualization designers must choose between simplicity, which can 
lead to overgeneralization and invisibilization, or better representativity of the population, which can result in 
hypervisibilization and \change{visual} clutter.
Minimizing the number of categories can considerably reduce the number of items to process and, thus, facilitate data exploration and decision-making. Representing the 288 ethnic groups of the UK census \cite{ONS:2022:census} can give a sense of representativeness of the population but make the visualization \change{less time-efficient 
} 
for tasks that require a lot of attention and carefulness. This trade-off between simplicity 
and better representation raises an important question: 
To what extent can better representation of people from marginalized groups reduce efficiency or lead to biased decisions that may harm the people represented? 

To be clear again, we are not advocating the invisibilization of groups in data visualizations. On the contrary, we want visualization researchers to invest time in understanding the effects of this trade-off on tasks as complex as decision-making. For example, it would be interesting to study tools that help decision-makers explore different disaggregation levels while remaining simple. One notable example is VisPilot \cite{Lee:2019:VisPilot}, which supports the exploration of subgroups while avoiding drill-down fallacies that lead to incomplete insights. 
It should be noted that 
many visualizations, in particular those designed for the general public, do not require the same type of \change{efficiency} 
as discussed above. Instead, designers should focus on other objectives, such as the representativity of the people visualized, understandability, or message reception. 

Few previous works in the visualization community have discussed the effect of grouping people \cite{Schwabish:2021:Harm,Baumer:2022:OfCourseItsPolitical}. Future studies can take inspiration from other research domains, such as the social sciences \cite{Herzog:2018:Invisibilization,Settles:2019:Hypervisibilize}, \change{to develop new methods to investigate how people feel harmed or represented when looking at visualizations under different levels of aggregation. }

\setlength{\picturewidth}{\columnwidth}
\begin{figure}[tb]
 \centering
\includegraphics[width=1\columnwidth]{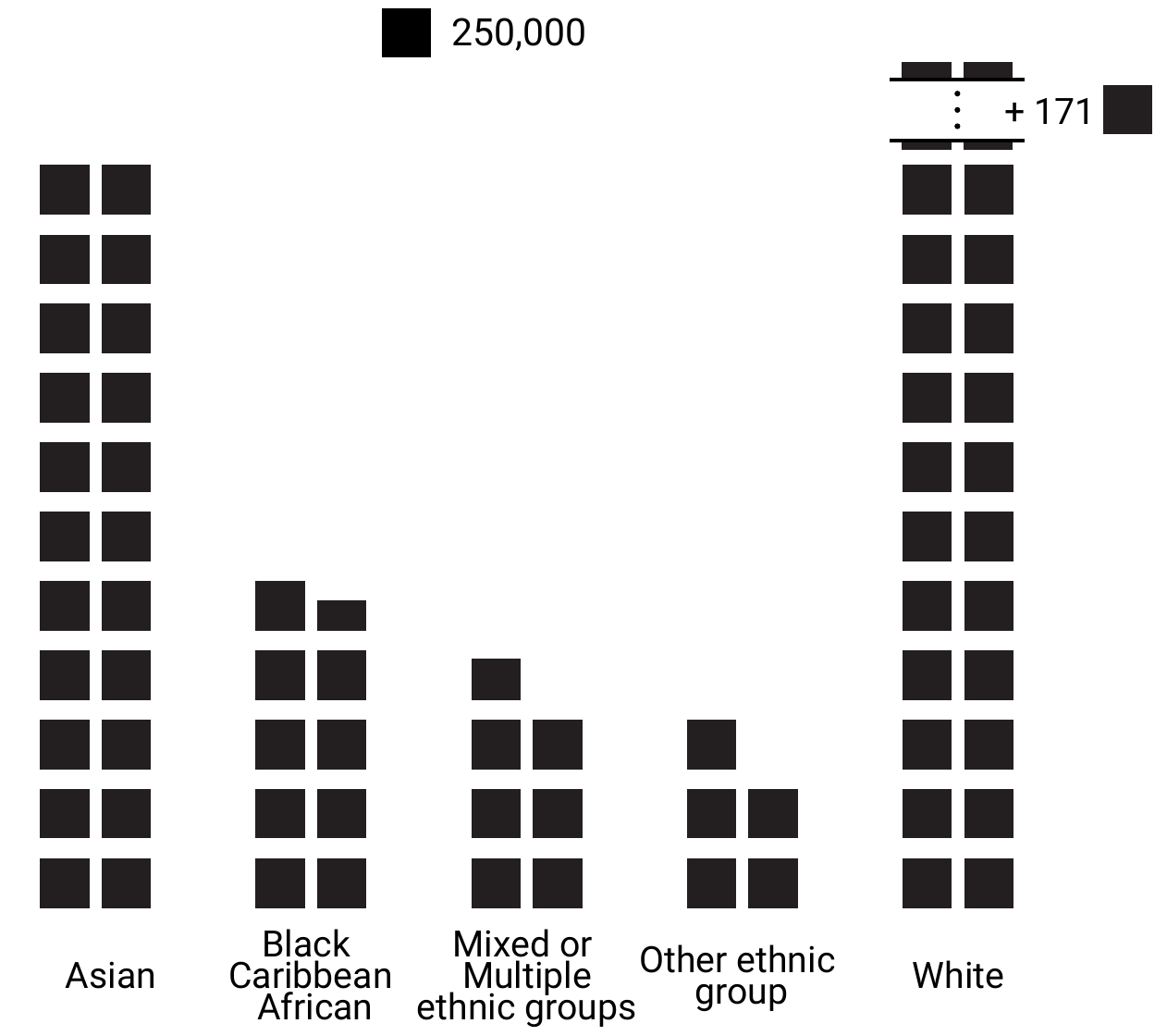}
 \caption{Unit-based visualization from the UK census \cite{ONS:2022:census}, where each full square represents 250,000 UK residents by ethnic group. While a unit-based visualization may be beneficial to show that each group is made up of individuals, \change{
 our choice of unit size doesn't allow full representation of the White group.} 
 }
 \label{fig:intermediatelevel}
 \vspace{-2em}
\end{figure}
\subsection{Inclusiveness of Different Representation Types}


Some visualization researchers have raised the question of how best to represent the individuals behind sociodemographic data. 
Recently, researchers have 
studied visual representations to show the diversity \cite{Dhawka:2023:Race} or disparity \cite{Holder:2023:Diversity} of people inside a group. Morais and colleagues defined three levels of \textit{granularity} to represent individuals \cite{Morais:2022:AnthropoDesignSpace}:
\begin{description}[\compact]
\item[\inlinevis{-1pt}{1em}{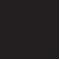} Low:] The marks aggregate all people from the same group to hide variability (see \autoref{fig:fivegroups} and \autoref{fig:twentygroups}).
\item[\inlinevis{-1pt}{1em}{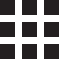} Intermediate:] The marks visualized represent a fixed number of people (greater than 1; see \autoref{fig:intermediatelevel})
\item [\inlinevis{-1pt}{1em}{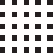} Maximum:] Each mark is associated with one individual.
\end{description}

This continuum poses interesting challenges for the visualization community. First, it is unclear how people represented in the data are affected by specific types of visualization or how they prefer to be represented. Second, it is unclear how a specific representation affects the stakeholders' understanding of the underlying data. 

Holder and Xiong \cite{Holder:2023:Diversity}, for example, showed that the type of representation mattered. The authors compared visualizations of low granularity (bar charts, dot plots) with those of higher levels of granularity (jitter plots, prediction intervals). The authors found low granularity visualizations 
increased stereotyped conclusions.

The representation of intermediate or maximum levels of granularity seems promising. However, visualizing individuals has multiple challenges, including the scalability of maximum granularity to a large dataset \cite{Morais:2022:AnthropoDesignSpace}, 
\change{and finding an adequate unit size, in particular when representing groups with large differences in population size. In \autoref{fig:intermediatelevel}, we decided on a unit size that allowed a more refined representation of smaller groups 
but 
made it difficult to fully represent the largest group within a suitable layout.}
Perhaps early work on focus+context visualization should be revisited in this context \cite{Tominski2017Lens}. 

A self-perception of the individual in data visualization can also be achieved with \textit{anthropographics} --- data visualizations that use human-shaped graphical representation. Researchers studied such visualizations to elicit more empathy for human rights \cite{Boy:2017:Antrhopographics} and promote prosocial behavior \cite{Morais:2021:Anthropographics}. 
Despite initial results that did not show clear benefits of anthropographics, this stream of research is growing. Notably, Dhawka \etal \cite{Dhawka:2023:Race} recently investigated the challenges 
of representing the diversity of people's races. \change{In particular, they identified the mapping between sociodemographic groups and visual channels as one that can lead to problematic choices, depending on whether or not the designer is aware of the social construction of groups and the stereotypes that can exist between the group and the visual channel. In the context of Virtual Reality, Ivanov \etal \cite{Ivanov2019Anthropohraphics} investigated individual representations of 3D human figures who died in mass shootings. 
The authors rendered human figures from six unique 
models as flat-grey silhouettes 
according to age and gender.} Anthropographics, as a new paradigm in visualization 
, requires further studies to understand its power, in particular, to promote diversity, equity, and inclusion.

Anthropographics per se requires grouping and visualizing people based on a physical characteristic shared by an entire group. In the ``Do No Harm Guide'' \cite{Schwabish:2021:Harm}, Schwabish and Xiong advised choosing unstereotyped icons to represent people. However, grouping people based on a physical characteristic requires that the entire group shares the physical characteristic and that this characteristic is not a topic of discrimination. Anthropographics raises many questions, including: To what extent is it even possible to represent people based on a physical characteristic that is not stereotyped? Also, to what extent does showing certain types of human-shaped icons lead to feelings of inclusion or exclusion?

\change{
\subsection{Summary}
As mentioned in Section 2, visualizations are not neutral. They can lead to biased decisions, the perpetuation of stereotypes, the reinforcement of discrimination, or the amplification of groups' marginalization.
Visualizing demographic data has the potential to cause great harm, thus requiring careful and responsible design choices.
Visualization designers can select what data they want to show, the visual marks mapped to each category, and what extra information to disclose. Such decisions 
place the designer in charge of how the visualization can be used or interpreted \cite{Correll:2019:Ethical} and, consequently, influencing the people visually represented. Designers must be aware that any design choice 
involves trade-offs between reducing potential harm and, for example,  
time efficiency, visual simplicity, or efforts to elicit empathy. 

We know little about the potential side effects of designing nonharmful visualizations. The set of open questions we raised focused on the trade-off between the use of visual stereotypes and efficiency, the trade-off between the inclusiveness of the marginalized group represented and simplicity, and the right level of individuation and perception to convey 
a sense of inclusion. But many other important questions remain open, including the efficiency of anthropographics or the scalability of unit-based visualizations. 

}
\section{Discussion}

At the beginning of this paper, we introduced protected attributes as important data dimensions to focus on in order to accurately represent individuals and groups of people and their characteristics. 
Overall, there is very limited visualization research on this topic, and until now, it has been mainly focused on three of these attributes: gender \cite{Muth:2018:alternative}, disability \cite{Barstow:2019:Symbols}, and ethnicity/race \cite{Dhawka:2023:Race}. We have not come across any dedicated discussion on the representation of religion, age, marital status, or any of the other attributes. The question is, why? Are some of these attributes less associated with visual stereotypes than others? For example, stereotyped colors come to mind immediately when we think of gender; 
conversely, fewer and weaker visual stereotypes may come to mind when considering age.

As researchers from both the US and Europe, we also are aware of the continental and national differences in calls for social changes broadly linked to diversity, equity, and inclusion. For example, a person's veteran status is a protected attribute in the US, while it is not in the UK or Canada. As such, studies on how visualizations of sociodemographic data are perceived should be a multi-cultural effort to study how and if perceptions are local.

\change{Including people from marginalized populations throughout the design process is an important step in mindful representations of such groups --- as suggested in recent works on demographic and nonharmful visualization \cite{Schwabish:2021:Harm,Dhawka:2023:Race,Dork:2013:Political,Baumer:2022:OfCourseItsPolitical,Barstow:2019:Symbols}. For example, Mack \etal \cite{Mack2023Avatar} interviewed 18 people with disabilities or related identities across two months to understand 
their perception, design choices, and use of avatars on online platforms. 
Respondents expressed the need to modify the avatar's disclosure of their disability according to context or environment. 
The choice of disclosing (or not) or aggregating (or not) 
personal data or attributes in visualizations should be made in consultation with the persons represented.

Visualization designers should also consider disclosing design
rationales and information about the people involved in the design process as an integral part of a visualization \cite{Burns2021WorkshopInvisible}. D\"{o}rk \etal \cite{Dork:2013:Political} 
highlight the importance of empowering the reader by providing background information about the data shown, the data hidden, and the intention of the visualization designer. Recently, Burns \etal \cite{Burns2022Invisible} showed that disclosing metadata, such as design rationales 
(for instance, why the designer mapped a specific visual variable to a category) or data sources (where the data comes from) can improve transparency and viewer's understanding.
}

Many topics related to sociodemographic data visualization require a cross-disciplinary approach. For instance, inclusiveness of data collection requires collaboration between scientists and institutions that collect data. Visualization researchers also have a role to play in designing innovative data collection methods. For example, Beischel and colleagues \cite{Beischel:2021:Collection} designed a visual tool to let participants express their sex and gender based on multiple dimensions.

\section{Conclusion}
Visualizing sociodemographic data requires designers to address many design challenges, from data collection to visualization
. The design of visualization is not neutral and can convey different messages with the same underlying data. Sometimes, these messages can harm people, in particular, people from marginalized groups, and we need to understand these potential harms better. Through a set of open questions, this paper aims to advance the discussion on how to design the least harmful visualization. 

Many of the questions we raise represent current considerations in the visualization community and in society. It should be noted that all the constructs that can harm people (invisibilization, hypervisibilization, stereotypes in representation, bias in decision-making) are socially and contextually dependent. Therefore, inclusive visualization practices may always be a moving target. When we base these visualizations on key principles, progress can be achieved, a point made well in the ``Do No Harm Guide'' \cite{Schwabish:2021:Harm}. 
We have an obligation to be careful and inclusive, but, more critically, additional research is needed to understand better how our design choices affect both the people represented and the people using our visualizations. 

\section*{Figure Credits}
We, as authors, state that all of our own figures are and remain under our own personal copyright, with permission to be used here. We also make them available under the Creative Commons Attribution 4.0 International (CC BY 4.0) license and share them at \href{https://osf.io/a2u9c/}{osf.io/a2u9c}

\section*{Acknowledgments} 
The authors wish to thank Stephanie Manuzak for proofreading the first version of the paper. This work was supported in part by the Action Exploratoire (AeX) EquityAnalytics from Inria.

\bibliographystyle{abbrv-doi-hyperref}

\bibliography{templateshort}
\end{document}